\documentstyle[11pt,psfig,newpasp,twoside]{article}
\def\edcomment#1{\iffalse\marginpar{\raggedright\sl#1\/}\else\relax\fi}
\reversemarginpar

\begin{document}
\title{The Binary Origins Of Hot Subdwarfs: New Radial 
Velocities}
\author{Rex A.\ Saffer}
\affil{Dept.\ of Astronomy \& Astrophysics, Villanova University, 
  Villanova, PA 19085}
\author{Elizabeth M.\ Green \& Timothy P. Bowers}
\affil{Steward Observatory, University of Arizona, Tucson, AZ 85721}

\begin{abstract}
Multiple radial velocity observations have been obtained for a large
sample of local field subdwarf B (sdB) stars over a period of two
years.\footnote{Observations reported in this paper were obtained at
the Multiple Mirror Telescope Observatory, a facility operated jointly
by the University of Arizona and the Smithsonian Institution.}  SdB
stars appear to fall into three distinct groups based on their
kinematic and spectroscopic properties.  The sdB's in Group~I have no
detectable spectral lines from a cool companion, and show only small
or insignificant velocity variations.  They represent about 35\% of
all sdB stars, which defines the upper limit for the total fraction of
non-binary sdB's.  Group~II sdB's are single-lined spectroscopic
binaries, comprising about 45\% of the sample.  Their spectra resemble
those in the first group, but they have significant or large velocity
variations and probable orbital periods on the order of a day.  The
third group contains the remaining $\sim 20$\% of sdB's, those showing
additional spectral lines from a cool (FGK) main sequence or subgiant
companion.  All Group~III sdB's have slowly varying or nearly constant
velocities, indicating periods of many months to several years.
Group~II sdB's are obvious post-common envelope systems, in strong
contrast to the wide binaries of Group~III.  Current data are
insufficient to rule out the possibility that some sdB's in Group~I
might be analogs of Group~III binaries, with undetectably faint
companions.  The clear division into three groups with such disparate
properties suggests very different evolutionary histories even though
the current physical states are essentially indistinguishable.
\end{abstract}


\section{Observations and Reductions}

We have obtained multiple precise velocities for more than 70 bright
subdwarf B stars, using the MMT Blue Channel spectrograph at 1\AA\
resolution from 4000--4930\AA.  Typically, each sdB star was observed
five to seven times over a 1 to 2 year period.  Most stars were
observed at intervals of 1 to 2 days, 1 to 2 months, 3 to 6 months,
and 12 to 18 months.  We were able to determine very precise
velocities, 1 $< \sigma_{\rm v} <$ 2~km~s$^{-1}$ in most cases,
primarily by exposing to a rather high S/N of about 70 to 100 for each
exposure and taking extraordinary care at every stage of the
reductions, but also due to a relatively long CCD (3Kx1K pixels) with
good cosmetics and readnoise.

The data were reduced with standard IRAF\footnote{IRAF is distributed
by the National Optical Astronomy Observatories, under cooperative
agreement with the National Science Foundation.} reduction routines
and analyzed with IRAF's radial velocity cross-correlation task
FXCOR, which uses the method of Tonry \& Davis (1979).  Our relatively
high resolution and high S/N allowed us to set the Fourier filter
parameters to cross-correlate on only the narrowest spectral features,
primarily helium and various weak metal lines, with only a small
contribution from the sharp cores of the Balmer lines.

We initially used a set of multiply-observed, bright, narrow-lined O
and B stars as radial velocity standards (Fekel \& Morse, as reported
by Stefanik \& Latham 1992).  Perversely, the only good
cross-correlation templates for sdB stars are the stars themselves.
The observations for individual sdB's were correlated against each
other, corrected to an approximate rest velocity with roughly the same
zero point as the OB standards, and combined into a preliminary
template.  The individual spectra were cross-correlated against the
combined spectrum and the process was iterated.  Final velocities for
each sdB were obtained from correlations with an extremely high S/N
super-template consisting of the sum of the ten most similar combined
sdB spectra (after iteratively readjusting the velocities of the
combined spectra to ensure a consistent zero point.)

A precise understanding of the velocity errors is crucial for
understanding the binary properties of sdB stars.  We conducted an
exhaustive Monte Carlo error analysis, following the procedures of
Pryor, Latham, \& Hazen (1988) and Armandroff, Olszewski, \& Pryor
(1995), to determine the proper scaling factors between the true
velocity errors and the relative values of VERR that were output by
FXCOR.  A comparison of our estimated errors for two stars in the
sample with well-determined, unique orbits, PG0941+280 and PG1101+249
(Green, in preparation), shows very good agreement.  The standard
deviations of the MMT residuals about the predicted velocity curve are
1.54 and 1.41~km~s$^{-1}$, respectively, and the Monte-Carlo
derived errors are 1.65 and 1.47 k/s.

A subset of sdB's, whose spectra show additional spectral lines from a
cool companion, required further processing.  They were
cross-correlated against super-templates of main sequence spectral
types from F6 to K5.  The best match, determined by the lowest FXCOR
error values, was used as a template to derive the velocity of the
cool companion.  The same super-template was also velocity-shifted,
scaled and subtracted from each individual composite spectrum to
recover the spectrum of the sdB component alone.  The resulting,
comparatively lower S/N sdB spectra were cross-correlated against the
best-matching high S/N ``pure'' sdB super-templates described above
to derive the sdB velocities.

\section{Discussion}

The non-composite sdB's comprise a kinematically unbiased sample that
naturally divides into two groups on the basis of the observed
velocity variations.  The 22 sdB's in our Group~I
show only small velocity variations, 1 $< \sigma_{\rm v} <$
4~km~s$^{-1}$.  The distribution of the statistical probabilities of
finding the observed variations is far from a normal gaussian.  There
are too many stars with bigger velocity variations than expected from
a random distribution with the known errors: 2/3 of the $\chi^2$
probabilities are less than 10\% and 1/3 are less than 0.1\%.
However, it is conceivable that pulsations could be responsible for
these small amplitude motions, leaving no compelling evidence either
for or against binarism in the stars in this group.

The 29 sdB's in Group~II have large velocity variations, with
$\sigma_{\rm v}$ distributed fairly uniformly from 7 to
91~km~s$^{-1}$.  For stars in this group, the velocity change
over a one day interval was typically a large fraction of the total
observed velocity range, indicating likely periods on the order of a
few hours to a few days.

Our Group~III consists of 21 composite spectrum sdB
stars\footnote{N.B.\ We observed additional sdB's already known to
have composite spectra, in order to increase the number of stars in
our sample in Group~III.}, for which we determined the velocities of
each component separately.  The velocity variations for both of the
individual components as well as the difference between the two
components were always small, less than a few km~s$^{-1}$.
Thus, the composite spectrum sdB's appear to have relatively
long periods of many months to several years.

The stars in all three groups appear to be bonafide sdB stars.
Figure~1 shows representative examples of combined spectra for
individual sdB's in each group.  The Group~I sdB's, in particular,
must be old disk stars with little halo contamination, given their
line-of-sight velocity dispersion of $\sigma_{\rm los}$ =
34.3~km~s$^{-1}$.

\bigskip
\centerline{\psfig{figure=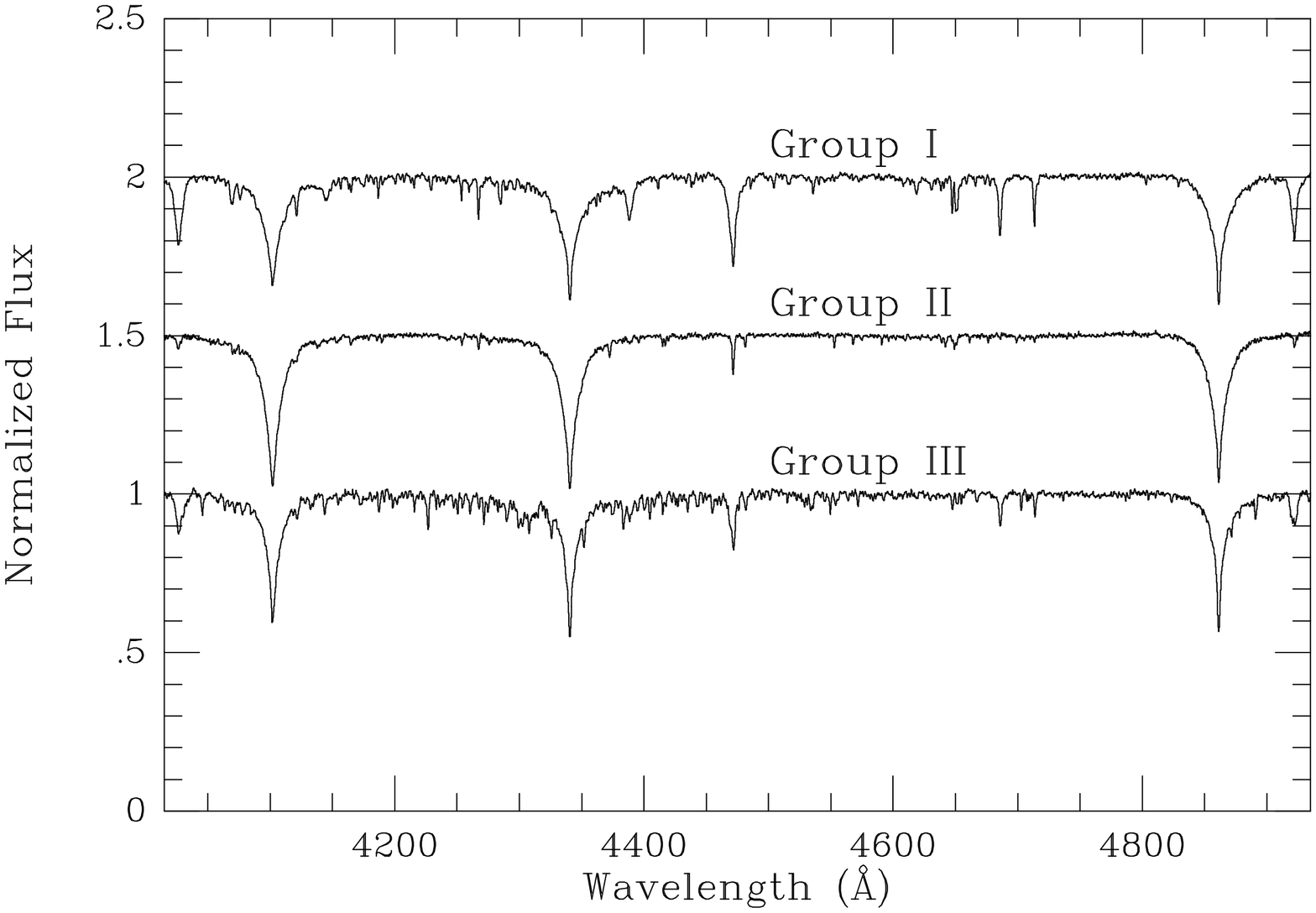,height=3.5in,angle=-90}}
\hspace*{0.3in} Fig.\ 1 -- Representative spectra of sdB's in Groups
I, II, and III.

\vfill\eject

\noindent
~~~~The reason for the dramatically different kinematic behavior
between Groups I and II is unknown.  There are no significant
differences between their distributions in the log $g$ {\it vs}
T$_{\rm eff}$ plane (Figure~2), as derived from Balmer line profile
fitting.  Both groups cover a surprisingly large range of helium and
metal line strengths, although there is a puzzling tendency for the
nearly-constant velocity Group~I stars to have stronger line
strengths, on average, than the short timescale velocity variables in
Group~II.  Group~I stars might also be somewhat brighter than those in
Group~II, although this is hardly a significant result (7\%) due to
the small number of stars in both groups.

\smallskip
\centerline{\psfig{figure=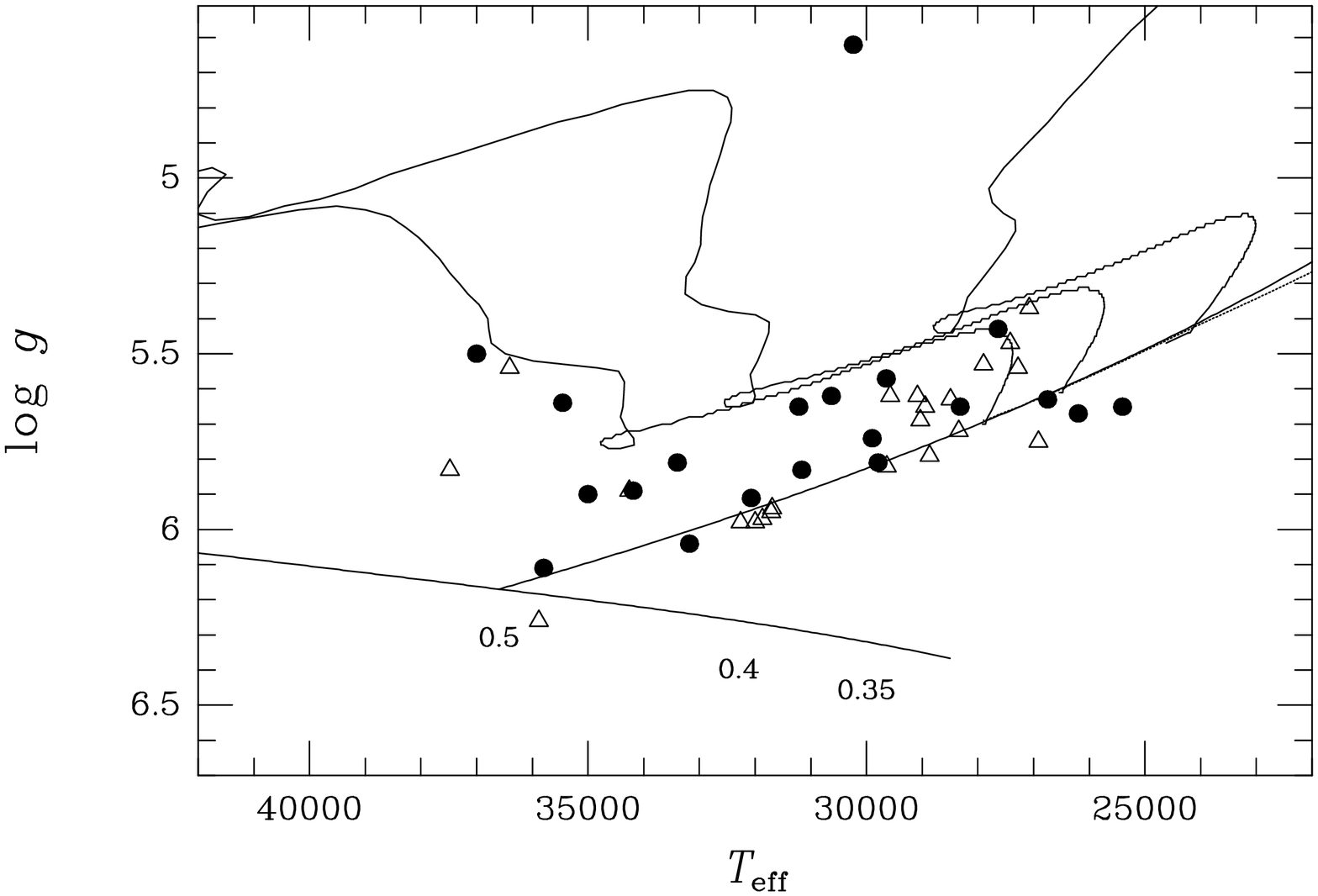,height=3.08in,angle=-90}}
\noindent Fig.\ 2 -- Existing log $g$ {\it vs} T$_{\rm eff}$ data for
stars in Groups~I (filled circles) and II (open triangles) relative to
evolutionary tracks and the He-burning MS.
\bigskip\medskip

We conclude that the clear division into three groups with such disparate
properties indicates quite different evolutionary histories.
The sdB's in Groups~I and III cannot have been through a common
envelope phase, whereas Group~II sdB's must all be post-common
envelope systems.  It is not yet clear whether any of
the Group~I sdB's are binaries or not.  Given our velocity accuracy,
no more than one sdB in Group~I is likely to be a face-on example of a
Group~II binary.  However, the small velocity variations of the Group~I
sdB's are similar to those of the sdB components of the majority of
the Group~III binaries, leaving open the possibility that at least
some of the Group~I stars could be wide binaries with undetected
companions.

\vfill

\end{document}